\documentclass{PoS}

\usepackage{graphicx}
\usepackage{cleveref}

\title{Search for neutrino emission in IceCube's archival data from the direction of IceCube alert events}

\ShortTitle{Self-triggered point source search}

\author{
The IceCube Collaboration\footnote{For collaboration list, see PoS(ICRC2019) 1177.}\\
{\itshape \href{http://icecube.wisc.edu/collaboration/authors/icrc19_icecube}{http://icecube.wisc.edu/collaboration/authors/icrc19\_icecube}}\\
E-mail: \email{mkarl@icecube.wisc.edu}
}

\abstract{

IceCube is a cubic-kilometer scale neutrino detector instrumenting a gigaton of ice at the geographic South Pole in Antarctica. On average, 8 track-like high-energy neutrino events with a high probability of being astrophysical are detected and published as alerts per year. The bright appearance of these events in the detector allow a precise pointing to their origins. This work presents a search for cosmic neutrino sources. The analysis uses high statistics archival IceCube neutrino-induced through-going muon samples to search for these sources in the vicinity of the incoming directions of the track-like high energy neutrino alert-events. The analysis searches for both steady sources emitting neutrinos over the entire uptime of IceCube, and transient sources that only temporarily produce neutrinos. This search will be applied to all historic alerts and will be automated for all future high energy track-like neutrino alerts.\\

\vspace{4mm}
{\bfseries Corresponding authors:}
\speaker{Martina Karl}$^{1, 2}$\\
{$^{1}$ \itshape Max Planck Institute for Physics}\\
{$^{2}$ \itshape Technical University of Munich}

}

\FullConference{36th International Cosmic Ray Conference -ICRC2019-\\
		July 24th - August 1st, 2019\\
		Madison, WI, U.S.A.}

\begin{document}

\section{Introduction}\label{sec:info}



\begin{figure}
    \centering
    \includegraphics[width=0.9\textwidth]{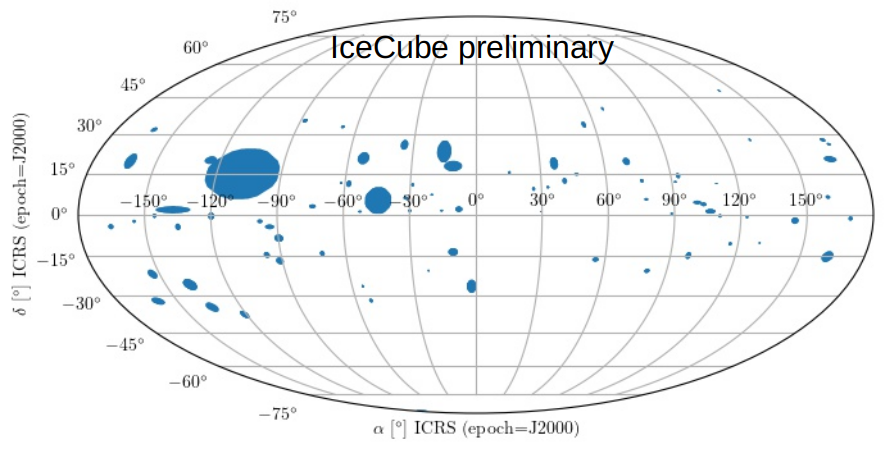}
    \caption{Skymap with the origin position of the issued alerts indicated as blue dots from August 2009 until March 2019. The size of the dot depicts the uncertainty region of the reconstruction, with a 90\% probability of the alert to be coming from inside the region.}
    \label{fig:skymap}
\end{figure}

The IceCube Neutrino Observatory \cite{Aartsen_2017} detects high energy neutrinos of astrophysical origin \cite{1242856}. The search for cosmic neutrino sources and counterparts is ongoing. Contrary to classical telescopes, IceCube has a field of view of the whole sky and is thus ideal to inform other telescopes of potentially interesting events. If a detected neutrino event has a high chance to be of astrophysical origin, the IceCube neutrino observatory sends alerts to other telescopes \cite{2017APh....92...30A}. These alerts trigger follow-up multi-messenger observations \cite{Kintscher_2016}. A map of the alerts can be found in \cref{fig:skymap}. This map shows all issued alerts until March 2019, and all events from before the alert system was implemented that fulfill the alert criteria, starting from August 2009. On the 22nd of September 2017, IceCube detected a neutrino (IceCube-170922A) with high energy and of astrophysical origin. The gamma-ray follow up observations found a flaring blazar at the origin of this event \cite{2018Sci...361.1378I}. Additionally, we searched for neutrino emission from that direction in archival IceCube data and found a neutrino flare between September 2014 and March 2015 \cite{2018Sci...361..147I}. 

This prompts the question whether there is neutrino emission coming from the direction of the remaining alert events. We look into 10.5 years of archival IceCube data and search for neutrino excesses coming from the alert origin. In this specific analysis, we search for a steady neutrino source. We use IceCube archival data from April 2008 until October 2018. A search for neutrino flares similar to what was observed for IceCube-170922A will follow this analysis in the near future.

\section{Analysis Method}\label{sec:refs}
We search for neutrinos coming from the direction of high energy tracks in IceCube. We use 10.5 years of IceCube data, comprising through-going muon tracks with energies $\gtrsim 100$~GeV \cite{Aartsen:2016oji}. The IceCube alerts provide the positions in the sky in which we are interested, similar to a catalog of expected neutrino sources. We use an unbinned likelihood formalism described in the following subsection. 

\subsection{Unbinned likelihood ratio}
We use an unbinned likelihood approach for our data analysis. The
likelihood function is given by the product over all events $i$ of the
probability density, obtained as a superposition of the densities for
signal ($S$) and background ($B$):
\begin{equation}
\mathcal{L} = \prod_{i} \left[ \frac{n_s}{N} S_i (\vec{x}_i , \sigma_i, E_i ; \vec{x}_s, \gamma) + \left( 1 - \frac{n_s}{N} \right) B_i (\delta _i, E_i) \right] ,
\end{equation}
where $n_s$ is the number of expected signal events in the detector, $N$ is the total number of events (signal + background) in the detector, $\vec{x}_s$ is the source position, and $\gamma$ is the source spectral index of an expected power law emission spectrum. For each event $i$, $\vec{x}_i$ denotes the reconstructed origin position, $\sigma _ i$ the one sigma uncertainty of the reconstructed position, $E_i$ the energy, and $\delta _i$ the reconstructed position's declination. The signal probability density functions $S$ are described by:

\begin{equation}
S(\vec{x}_i , \sigma_i, E_i ; \vec{x}_s, \gamma) = S_{spatial} \cdot S_{energy} ,
\end{equation}

\begin{equation}
S_i (\vec{x}_i , \sigma_i, E_i ; \vec{x}_s, \gamma) = S_i (\vec{x}_i, \sigma_i ; \vec{x}_s) \cdot \varepsilon_{S} (E_i ; \delta _i , \gamma ) = \frac{1}{2 \pi \sigma_i ^2} \exp \left( - \frac{- | \vec{x}_i - \vec{x}_s | ^2}{2 \sigma _i ^2} \right) \cdot \varepsilon _{S} (E_i ; \delta _i, \gamma ) ,
\end{equation}
where the last factor, the energy factor $\epsilon _{s}$ is the probability density function for a signal event of Energy $E_i$ given its declination $\delta _i$ and the source spectral index $\gamma$. 
This energy factor is calculated from Monte Carlo data \cite{Braun:2008bg}.

Similarly, the background probability density functions $B$ are expressed as:

\begin{equation}
B(\vec{x}_i, E_i) =B_{spatial} \cdot B_{energy} ,
\end{equation}

\begin{equation}
B \left( \vec{x}_i, E_i \right) = B_i (\vec{x}_i) \cdot \varepsilon _{B} (E_i; \delta _i) = \frac{1}{2 \pi} \cdot P(\delta _i) \cdot \varepsilon _{B}(E_i; \delta _i) ,
\end{equation}
where we assume uniformity over right ascension\footnote{The IceCube detector is directly located at the Geographic South Pole. Because of earth rotation, we see the same background everywhere in right ascension for integration times longer than a few days.} and only a spatial dependency over declination $\delta _i$. Here again, the energy term $\varepsilon _{B}$ describes the probability density function for a background event with energy $E_i$ at declination $\delta _i$.

For the test statistic ($TS$) we take the likelihood ratio of the null-hypothesis (background only, $n_s = 0$) to the best fit of the signal hypothesis $n_s > 0$. The ratio can then be expressed in the following way: 

\begin{equation}
TS = -2\log \left[\frac{P(Data|H_0)}{P(Data|H_S)} \right] ,
\end{equation}

\begin{equation}
TS = -2~\log \left[\frac{\mathcal{L}(n_s=0)}{\mathcal{L}(n_s=\hat{n}_s)} \right] =
                       2 ~ \sum_{i} \log \left[\frac{\hat{n}_s}{N_{obs}}\left(\frac{S_i}{B_i}-1\right) +1\right].
\end{equation}

\subsection{Position fit}
\begin{figure}
    \centering
    \begin{minipage}{0.49\textwidth}
        \includegraphics[width=\textwidth]{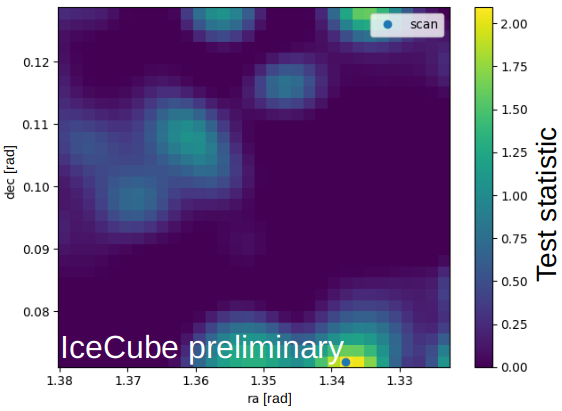}
    \end{minipage}
    \begin{minipage}{0.49\textwidth}
        \includegraphics[width=\textwidth]{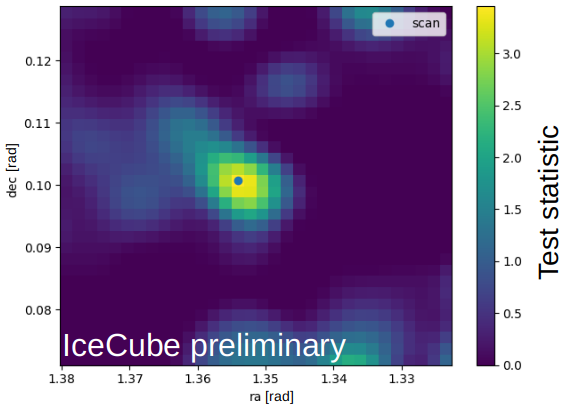}
    \end{minipage}
    \caption{Map of the test statistic values for a random background distribution (left) and a region with injected signal (right). The region is divided into steps with $0.1 ^{\circ}$ spacing. At each step the best test statistic value is determined by fitting the best mean number of signal events and spectral index. The position yielding the highest test statistic value is then considered to be the point source position, here indicated as the blue dot, labeled "scan" as for the result of the region scan. The left plot shows a random background and the scan finds the position with the highest background fluctuation. In the right plot, there are 10 signal events injected at the center of the plot and the scan finds the position where the events were injected.}
    \label{fig:pos_scan_example}
\end{figure}

We treat the alert events in IceCube as a source catalog. Contrary to an astrophysical source catalog, we do not get a precise, point-like position for our sources. However, we do not expect our sources to be extended. For example, the counterpart to IceCube-172209A is the Blazar TXS0506+056 with a diameter in sub-arc minute region \cite{2010AJ....139.1713C}. With IceCube, sub-degree, however not sub-arc minute, resolution can be achieved. Thus, objects with arc minute extension or lower can be considered as point-like sources. We want to find the point-like source in the origin uncertainty region of the issued alerts. Therefore, we scan the position using a grid with $0.1 ^{\circ}$ spacing. At each point we maximize the test statistic value by fitting the number of signal events and source spectral index. In the end, we take the grid point with the highest test statistic value as our source position. This is illustrated in \cref{fig:pos_scan_example}, in which the position with the best test statistic value is indicated by "scan". Based on the expected distribution of the best-fit test statistic we determine a p-value for each of the 81 alerts.

\subsection{Individual fit}
There are two general scenarios of continuous sources that we want to cover with this analysis. The first case is that we have few strong sources. In order to cover this case, we look at each source individually and choose the most significant one as our result. This method searches for point sources, but with a method different from that of the regular IceCube untriggered all-sky search. Therefore a trial factor of close to two would apply if those results were combined.

\subsection{Stacking analysis}
The second source scenario is that we have many sources, whereas each on its own emits a relatively low flux so we could miss these sources in the individual case. With the stacking, we view the alerts as representing a population of sources but not as single source candidates. For this, the test statistic is the sum of all the individual test statistic values of all alerts $k$: 
\begin{equation}
    TS_{stacked} = \sum _{k} TS_{k} .
\end{equation}

\section{Results}
\begin{figure}
    \centering
        \includegraphics[width=0.9\textwidth]{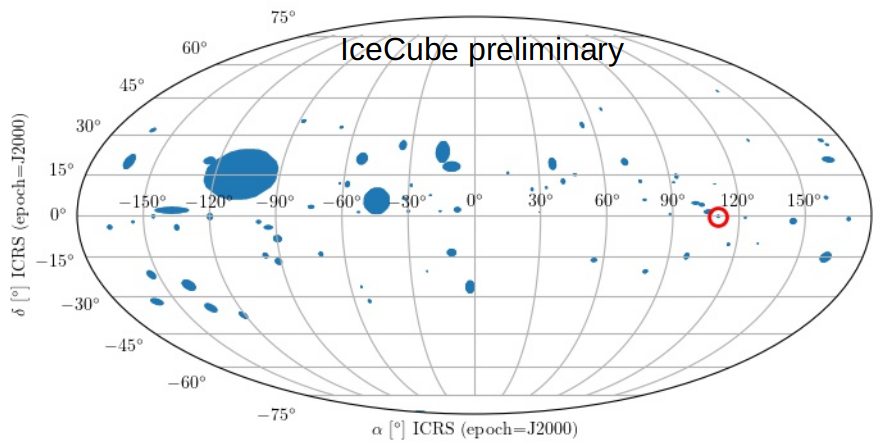}
    \caption{Skymap with all the alerts. The most significant alert is marked with a red circle. }
    \label{fig:indiv_result_map}
\end{figure}

We present the preliminary results of our time integrated search. For the individual source hypothesis, we find a best p-value of 0.83. Thus, our result is compatible with background. The location of the most significant alert is indicated with a red circle in \cref{fig:indiv_result_map}. The alert is close to the galactic plane.

\begin{figure}
    \centering
    \includegraphics[width=0.8\textwidth]{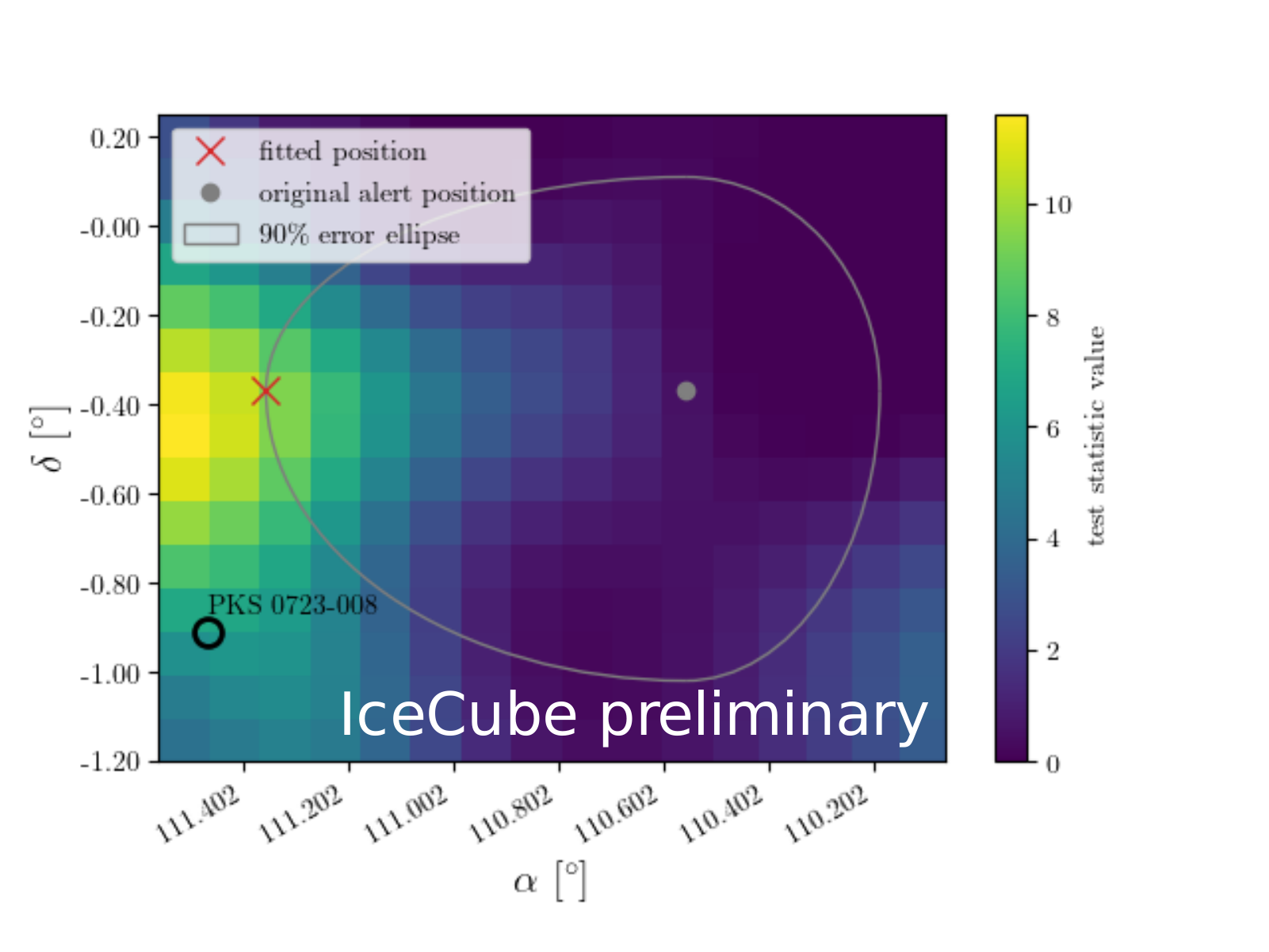}
    \caption{Scanned uncertainty region of the most significant alert. The x-axis shows the right ascension in degrees, the y-axis the declination in degrees. The 90\% error region of the reconstructed alert position is depicted as a grey circle, the original alert position is marked as a grey dot. The position with the highest test statistic value is indicated with a red cross. We only consider values within the 90\% error contour of the alert. At a distance of $\sim 0.33$~arc~minutes is the blazar PKS~0723-008 located \cite{2015AJ....150...58F}.}
    \label{fig:indiv_result_scan}
\end{figure}

We show the test statistics map of the scanned uncertainty region of the most significant alert in \cref{fig:indiv_result_scan}. We only consider values within the 90\% error contour of the reconstructed alert direction. The position yielding the highest test statistic value is marked with a red cross. A blazar is located $\sim 0.33$~arc~minutes from the fitted source position and is indicated with a black circle \cite{2015AJ....150...58F}. 

For the stacking, we get a p-value of 0.856. The scenario of a source population is also compatible with background. Additionally, we are working on a follow up analysis looking for time-varying emission. Results for the transient source search will follow soon. 



\bibliographystyle{ICRC}
\bibliography{references}

%

\end{document}